# Andreev bound states in superconductor/ferromagnet point contact Andreev reflection spectra

K. A. Yates

*Physics Department, The Blackett Laboratory, Imperial College London, SW7 2AZ, United Kingdom,*

D. Prabhakaran

*Department of Physics, Clarendon Laboratory, University of Oxford, Park Road, Oxford OX1 3PU, UK*

M. Egilmez*, J. W. A. Robinson

*Department of Materials Science and Metallurgy, University of Cambridge, 27 Charles Babbage Road, Cambridge CB3 0FS, United Kingdom*

L.F. Cohen

*Physics Department, The Blackett Laboratory, Imperial College London, SW7 2AZ, United Kingdom,*

*Current Address: Department of Physics, American University of Sharjah, Sharjah 26666, United Arab Emirates

As charge carriers traverse a single superconductor ferromagnet interface they experience an additional spin-dependent phase angle which results in spin mixing and the formation of a bound state called the Andreev Bound State. This state is an essential component in the generation of long range spin triplet proximity induced superconductivity and yet the factors controlling the degree of spin mixing and the formation of the bound state remain elusive. Here we demonstrate that point contact Andreev reflection can be used to detect the bound state and extract the resulting spin mixing angle. By examining spectra taken from $La_{1.15}Sr_{1.85}Mn_2O_7$ single crystal – Pb junctions, together with a compilation of literature data on highly spin polarised systems, we show that the existence of the Andreev Bound State both resolves a number of long standing controversies in the Andreev literature as well as defining a route to quantify the strength of spin mixing at superconductor-ferromagnet interfaces. Intriguingly we find that for these high transparency junctions, the spin mixing angle appears to take a relatively narrow range of values across all the samples studied. The ferromagnets we have chosen to study share a common property in terms of their spin arrangement, and our observations may point to the importance of this property in determining the spin mixing angle under these circumstances.

## I. INTRODUCTION

Andreev reflection spectroscopy has been widely employed to determine transport spin polarisation in ferromagnets. The seminal 1998 paper by Soulen *et al.*, [1] opened up the field of point contact Andreev spectroscopy (*PCAR*) demonstrating its important application for determining the degree to which electrical currents carry an excess of one direction of spin, the so-called transport spin polarisation (*P*). The use of PCAR to determine *P*, is based on the premise that at an interface between an s-wave superconductor and a normal metal, and at voltages below the superconducting gap voltage, an electron with one spin direction (e.g. spin up), couples with an electron with the opposite spin (i.e. spin down) to pass into the superconductor as a Cooper pair, requiring the retro-reflection of a hole of the opposite spin (i.e. spin down) back into the normal metal. This leads to a doubling of the conductance at subgap voltages (for reviews see [2,3]). As the *P* of the metal increases, the probability of retro-reflection for the holes decreases and consequently the Andreev reflection process is suppressed and the sub-gap conductance decreases. The degree to which the Andreev signature (i.e. the doubling of the differential conductance at bias voltages less than the superconducting energy gap) is suppressed is an indication of the degree of *P* [1]. The ever growing interest in spin-dependent devices, where the polarisation of the transport carrier has significant importance in device function (e.g. spintronics, spin Seebeck, magnonics) means that there has been enormous interest in a technique that promised to provide a definitive determination of *P* of new materials and heterostructures. The original theoretical description of Andreev conductance across a superconductor-normal metal (SN) contact, the "BTK" model [4,5] was adapted elegantly to incorporate the parameter of transport spin [6]. The accessibility of the Mazin-BTK (M-BTK) method resulted in its widespread use [7,8,9,10,11,12,13,14,15]. As a result of this large body of work, it became clear that certain spectroscopic features lay outside the scope of the original model, leading to the development of further descriptions of the interface properties between a superconductor and a ferromagnet (SF) [16, 17, 18].

Around the same time, the emerging interest in SFS junctions, or π junctions, was evolving a parallel and detailed understanding of the influence of the ferromagnetic exchange field on the superconducting phase difference of Josephson junctions [19, 20, 21, 22, 23,24]. This understanding included the description of bound states existing in the junction itself, the so-called Andreev Bound States (ABS) [22,24,25]. The formation of the ABS is also a critical component in the creation of a long-range spin triplet proximity effect (LRSTPE) [26,27] for which there is now conclusive evidence in the literature [28,29,30,31,32,33,34,35]. In the LRSTPE, spin singlet Cooper pairs are converted to long-lived odd frequency spin triplet states in the ferromagnet by virtue of a two step process, first spin mixing resulting from the introduction of a spin dependent phase angle between quasiparticles at the interface [36], followed by spin flip scattering in the presence of inhomogeneous spin order at that interface (see [37] for a review). As a component of the developing description of LRSTPE, the theory also predicted that ABS must exist at each single SF constriction contact [38,39]. Although scanning tunnelling microscopy (STM) measurements have indicated a change to the density of states of both the F or the S layers that is consistent with the theory of LRSTPE generation [40, 41, 42, 43], as yet only one clear experimental observation of ABS at single interfaces exists [44]. In order to determine whether ABS are observed routinely in conventional point contact spectra albeit in a form that allows them to be easily overlooked, we analyse the sub-gap conductance of Pb point contacts formed onto $La_{1.15}Sr_{1.85}Mn_2O_7$ (LSMO), and re-examine data taken on highly spin polarised materials found in the literature [7,14,15,45].

In part II of the paper, the fitting of Andreev reflection spectra is reviewed. In part III the effects of ABS on the measured Andreev reflection spectra are considered. Part IV considers indirect experimental evidence for ABS while part V explores direct measurements of ABS. The final part of the paper summarises our findings.

## II. TRADITIONAL OBSTACLES TO THE DETERMINATION OF *P*

The success of the BTK model was the incorporation of the concept of a dirty interface using the delta function parameter Z. Z reduces the transparency of the Andreev contact [4,5]. For low Z the barrier is transparent to conduction electrons and conduction into the superconductor at sub-gap bias occurs via Andreev reflection. At high Z, the contact is tunnel-like. For intermediate values of Z conduction is modelled as involving contributions from both Andreev reflection and tunnelling currents. The BTK model gave the conductance as a function of Z for unpolarised contacts [4,5] The M-BTK model incorporated both *P* and Z using the Landau Buttiker formulism [6]. The focus of early questions, once the M-BTK model had been established, related to the meaning and nature of Z and the physics that might be contained within it [13,46,47]. As both *P* and *Z* act to suppress the conductance and as, ultimately the `measurement' of *P* via Andreev spectroscopy involves *fitting* a conductance curve, it was recognised that the effects of *Z* and *P* could become indistinguishable, particularly in the presence of thermal (and non-thermal) broadening [10,12].This *P-Z* compensation effect however did not initially prove problematic as the community quickly developed methods of fitting the data so as to reduce the *P-Z* ambiguity [12,13].

Although Z had initially been introduced to account for interface transparency, it was shown to have additional physical meaning [2,5,47]. The effect of Fermi velocity mismatch at the S-F contact provided a contribution to Z [5,18,45] resulting in a predicted lower bound for Z ($Z_{eff}$) for each of the spin bands [18] showing that *Z* = 0 would be unphysical. The 'Z=0 problem' is a feature present in the PCAR results of many highly polarised materials [48], particularly the fully spin polarised $CrO_2$ [45]. In addition to *P* and Z, other parameters were introduced into the fit routine, including the superconducting energy gap parameter, $\Delta$, and a spectral broadening parameter, $\omega$ (or $\Gamma$) [12] (equivalently either an effective temperature [8,10] or spreading resistance [13]). Physical justifications given for the introduction of these parameters included the observation that $\Delta$ may not take the bulk value at a point contact under mechanical stress [12]. Many influences could result in spectral broadening, including a spread of interface parameters across the interface, inelastic quasi-particle scattering and inhomogeneous disorder [47,49]. Additionally a dependence of *P* on Z was widely observed [13, 46,48,50]. Although theories were proposed for the origin of this dependence [46], no unique mechanism has been agreed [13]. Nonetheless, contacts showing high polarisation were repeatedly observed to have low values of *Z* that are unphysical and the method of extrapolation to zero *Z* as a determinant of bulk spin polarisation became quite widespread [48,50].

The highly spin polarised material $CrO_2$, demonstrates both the *P(Z)* issue and the 'Z=0 problem', Fig. 1. For this 100% spin polarised material, the highest polarisation has often been determined from spectra that, when fitted, resulted in very low, or zero, Z values [7,38,45,48,51]. For the particular case of $CrO_2$ interrogated with a Pb contact, the Fermi velocity mismatch results in a minimum Z of ~ 0.26 [45]. As Figure 1 shows, those contacts with an apparent extracted *P* > 85%, all have an extracted *Z* parameter lying below this minimum. In addition to the *P(Z)* dependence, plotting the other extracted parameters against *Z* reveals that contacts showing a high *P* are often associated with a $\Delta$ value that is depressed [45], figure 1b. Although significant non-thermal broadening also exists in the spectra shown in figure 1c, there is no discernable correlation between the degree of suppression of $\Delta$ and the magnitude of the spectral broadening, from which one can rule out inelastic scattering as a cause of Δ suppression [45]. Data taken from the literature fitted using a series resistance rather than a broadening parameter showed a similar trend for $CrO_2$ (fig 1c.) [38]. Similar results showing depressed Δ associated with spectra showing enhanced spectral broadening were also observed in (Ga,Mn)As [8].

Now let us review the existence of states in the gap. There are a number of special cases that produce enhancement of the sub-gap conductance. A magnetic scattering layer at some distance within the normal metal from the S-N interface produces sub-gap states (contributing to the conductance) caused by multiple carrier reflections between the superconductor, the normal metal and the magnetic scattering layer [52]. Similarly, interference effects are expected in simple SF interfaces if the ferromagnet is a thin layer [3,53]. Both of these cases are very specific states of the sample and easily identified.

Joule heating in the point contact area can produce zero bias anomalies if the contact is not in the ballistic limit (where the ballistic limit is defined as the contact diameter being less than both the mean free path in the ferromagnet and the superconducting coherence length [2,5]). However, this can be avoided by careful measurement and so should not be the cause of enhanced sub-gap conductance that has been observed systematically in $CrO_2$ [45] and (Ga,Mn)As [8].

There are other phenomena that have been predicted to show enhanced sub-gap conductance, such as a resonant enhanced proximity effect [19], or magnon assisted sub-gap transport [54]. These are more difficult to avoid and therefore cannot be categorically excluded as an explanation of the behaviour in Figure 1. However, despite reports of enhanced sub-gap conductance due to the proximity effect in superconductor-semiconductor structures [55,56], the effect of these predicted proximity issues in SF structures was assumed to be within experimental error (which is considerable for the point contact technique [12]), or controllable if the contact size was kept in the true ballistic limit [3,16,18]. Indeed, measurements of *P* via the Andreev technique for moderately spin polarised materials without accounting for these anomalous effects, gave results consistent with those of other techniques [12,18]. These studies motivate us to re-examine whether spin mixing effects could be responsible for the enhanced inelastic scattering observed in (Ga,Mn)As [8], $CrO_2$ [45] and Pt [18].

III. SPIN-DEPENDENT INTERFACTIAL SCATTERING

The realisation that spin orientation could result in phase differences between carriers as they traverse the SF interface [38,39] and not just different transmission probabilities [16,18,57,58], led to the appreciation that ABS may be an intrinsic feature of PCAR conductance spectra [26,38,39]. The models that had been developed in order to understand superconducting π junctions [20,24] were extended to single SF point contact interfaces [21,38,39,59,60,61]. These models introduced a spin mixing angle, θ, associated with the phase difference picked up by the spins as they cross the interface. The spin mixing angle can significantly affect the sub-gap conductance and in particular, should result in the appearance of ABS [39,59]. The spin mixing angle additionally lowers the intensity of the coherence peaks at the superconducting gap voltage, and shifts the spectral weight of this contribution to sub-gap voltages via the ABS [62].

It is this effect which may be of relevance to the 'Z=0 problem'. As the effect of ABS is to increase the sub-gap conductance without increasing the conductance at the gap voltage (i.e. without contributing to the coherence peaks associated with finite Z) then, were such a spectrum fitted without accounting for these ABS, the extracted fitting parameters would indicate either an artificially lower *P* or Z and/or include a significantly enhanced non-thermal broadening term to explain the finite states in the gap [8].

The differences between the M-BTK and the spin mixing model (SMM) can be illustrated by considering the limiting case of 100% spin polarisation, such as expected for $CrO_2$ [45]. For the M-BTK model, the absence of one spin species leads directly to a complete suppression of $G_0$ (fig. 2a). A finite conductance at zero bias for a 100% spin polarised material can only be achieved, within the M-BTK model, in the presence of finite broadening of the quasi-particle density of states, such as exists at finite temperature or through inelastic scattering. It is the sum of these broadening parameters that determines the temperature evolution of $G_0$ [47] and particularly it is these features that would be responsible for the `flattening off' of the $G_0(T)$ at low temperatures within the M-BTK model [38]. The case is entirely different within the SMM. Within this second model, each spin picks up a phase as it traverses the SF boundary. The phase difference allows fully spin polarised materials to form a Cooper pair of the triplet form ($|\uparrow\downarrow\rangle + |\downarrow\uparrow\rangle$). These Cooper pairs are then able to travel from S into F, that is, the Andreev process is not completely suppressed and the sub-gap conductance is no longer, necessarily, zero. This spin mixing is also the first step in the generation of the LRSTPE where a further spin flip process transforms these triplet Cooper pairs into the forms ($|\uparrow\uparrow\rangle$) and ($|\downarrow\downarrow\rangle$) [27,37]. ABS form at finite voltages in the PCAR spectra determined by the spin mixing angle, $\theta$ [24,39]:

$$\epsilon_{pole} = \pm\Delta\cos\left(\frac{\theta}{2}\right)$$

The ABS features are broadened at finite temperature and so contribute smoothly to $G_0$ (fig. 2b). As the temperature is reduced (and the thermal broadening decreases), the ABS become more clearly defined and the zero bias conductance reduces to its highly polarised value of zero. Although a zero bias ABS can appear, it is a special case of $\theta = \pi$, and the geometry of the contact, barrier and the Fermi surfaces of S and F combine to make such high values of $\theta$ unlikely [39]. For models that consider the effect of phase on spin transmission in PCAR spectra, it is expected that $G_0 = 0$ at T = 0 [39,60].

IV. STATES IN THE GAP AS DETERMINED BY THE TEMPERATURE DEPENDENCE OF $G_0/G_N$

Even in the presence of a continuum of ABS, or with ABS below the detection limit of the technique used, the behaviour of the $G_0/G_n$ curve, as described by Löfwander [38], should be relevant as a test of the SMM or M-BTK models, ie. as an indication as to whether ABS are intrinsic to PCAR spectra. With zero non-thermal broadening, the M-BTK model predicts an exponential-like decrease of the conductance with temperature for P = 100% [38] (Fig. 2(a)). In contrast, the SMM with a spin mixing angle of $\theta = \pi/2$ shows a slower decrease of $G_0$ with temperature (Fig. 2(b)) [38]. A moderate amount of non-thermal broadening however, makes the predictions of the two models harder to distinguish [63]. In Figure 2a the impact on the temperature dependent conductance of introducing additional broadening to the M-BTK model (*P* = 100%, *Z* = 0.1, $\Delta = \Delta(T)$ taken from [64]) is shown explicitly. From this we learn that in order to distinguish between models, measurements taken at T << $0.5T_c$ are required.

Figure 2(b) shows $G_0/G_n$ for $CrO_2$ films grown onto $TiO_2$ (blue data) [63], $Al_2O_3$ (dark yellow data) [45] and LSMO single crystals. Also shown are $G_0/G_n$ data extracted from spectra in the literature for $CrO_2$ grown onto $TiO_2$ (blue data) [15], the equally highly spin polarized $La_{0.7}Sr_{0.3}MnO_3$ (LSMO3) [7] and $HgCr_2Se_4$ [65]. There is a clear clustering of the data. For $CrO_2$ none of the $G_0/G_n$ (T) curves fall on that expected for the M-BTK limit of 100% polarisation. The data on LCMO and on $HgCr_2Se_4$ is taken to lower temperature [7,65] and also appears inconsistent with the M-BTK prediction for a fully spin polarised material. As shown in figure 2a, intriguingly the data for $HgCr_2Se_4$ particularly does not fit within a M-BTK model even with additional (temperature independent) non-thermal broadening.

V. SEARCH FOR ABS IN PCAR $d^2I/dV^2$

As yet only the study by Hübler et al., reports observation of ABS-like features in a number of Al-$Al_2O_3$-Fe tunnel junctions studied at 50mK [44]. The data from Hübler et al., suggest that in the tunnelling regime (low interface transmission), the absolute magnitude of the observed ABS is small requiring low temperatures for detection. The authors suggest that ABS features appeared in the conductance curves only in spatial regions where the tunnelling barrier was sufficiently thin. In this regime they find that the spin mixing angle takes a broad range of values reflecting its sensitivity to interface properties. In contrast, the transmission probability in PCAR spectra should be considerably higher than in a tunnelling measurement [39, 44], allowing detection at higher measurement temperature. It is therefore reasonable to search for symmetric 'bumps' in the point contact conductance spectra that correlate with a dip in $d^2I/dV^2$ in the positive voltage and a peak in the $d^2I/dV^2$ in the negative voltage.

A series of conductance spectra taken on $La_{1.15}Sr_{1.85}Mn_2O_7$ (LSMO) single crystals were examined for evidence of ABS. The crystals were grown by an optical floating zone imaging furnace under a mixed oxygen/argon atmosphere pressurized to approximately $6-8 \times 10^5$ Pa, and phase identification, purity and structural quality was determined by X-ray diffraction and electron probe microanalysis, as described in detail elsewhere [66]. Point contact measurements were taken using mechanically sharpened Pb tips ($T_c$ = 7.2K) using a differential screw mechanism to slowly bring the tip into contact with the sample in a liquid helium dewar [12]. Once a contact was established, spectra were taken such that either the pressure on the contact was varied (contact resistance was changed) or the temperature of the contact was slowly increased. Where data was fitted, the $\chi^2(P)$ technique was used whereby the spectra are fitted for $\Delta$, Z and $\omega$ for each value of $P_{trial}$ [12]. The value of $P$ is then the minimum in the $\chi^2(P_{trial})$ function as described in ref [12]. The second derivative was scrutinised for a symmetric peak/dip signature in the -/+ voltage of the $d^2I/dV^2$. An example of a spectrum and its derivative is shown in Fig. 3(b). This composition is expected to be close to a half metal [67,68,69] and indeed fitting the spectra with the M-BTK model indicated very high values (though not 100%) for the polarisation. The polarisation determined from the M-BTK model is listed for four contacts in table I, while the fit to the data for one contact is shown in Fig. 3(a). In order to eliminate random noise from what may be a real signature of an ABS, the following steps were taken:

1) Peak/dip features in $d^2I/dV^2$ were identified as potential ABS if and only if they occurred symmetrically with respect to the bias voltage.
2) Noise can occur on both the conductance and the voltage bias measurement, but ABS will only appear in the conductance. Consequently, the differential conductance data was plotted as a function of acquisition order (effectively the 'time' of measurement). The derivative of this conductance data was examined for peak/dip features symmetric about the zero bias dip in the conductance. Features found from the derivative of this conductance were then compared to the features that had occurred in the $d^2I/dV^2$. Features that appeared in both derivatives were identified as potential ABS features. This ensured that the features being examined were in the conductance part of the spectrum and not noise on the voltage reading.
3) The spin mixing angles determined using equation 1 were compared with the spin mixing angle extracted from the IV measurements.

These are strict criteria but imposing them facilitates the extraction of only symmetric ABS as predicted by the SMM [39].

The aim of our experiments was to obtain an examination of the evolution of θ for a particular point contact when either pressure (changing the interface resistance and *Z*) or temperature was varied. Nineteen spectra were obtained in total on LSMO. Sixteen spectra were taken as a function of contact resistance for several contacts while one series of three contacts were taken as a function of temperature. After analysis, eleven spectra showed features with symmetric peak/dip features in the $d^2I/dV^2$ verses voltage. When the contacts were further analysed with step 2, eight spectra showed near symmetric bumps in the derivative vs row number, but only four spectra showed features in the conductance at approximately the same voltage as in the $d^2I/dV^2$. Of these, three spectra were of a contact undergoing systematic variation (with contact resistance). Values of θ estimated from the conductance spectra (by applying equation 1) are listed in table I. Δ was extracted from each dI/dV spectrum using the position of the coherence 'peaks'. No systematic change in the extracted θ value with contact resistance was observed. It is however interesting to note that there is a clustering of θ extracted from the $d^2I/dV^2$. Although measured in an entirely independent way, this is consistent with the clustering of data shown in figure 2. In combination these observations suggest that in the junctions we have examined here, the average spin mixing angle appears to take a narrow range of values grouped around 0.5π. Our observations also suggests that the elusive ABS may indeed be evident in many low temperature Andreev point contact conductance spectra.

The excess current can in principle be used as a third independent check of θ. The excess current is defined as $IR_n/\Delta$. In our case $R_n$ was determined from the gradient of the IV at |5-15| mV and Δ was either determined from the M-BTK fit or assumed to take the bulk value as assumed in ref [38]. The values of excess current were compared with those values given in figure 4 of ref [38]. Although it is not possible to extract exact values of θ in this way given that in that paper it is assumed that P = 100%, nevertheless by using Z=0 as an upper limit, $\theta_{max}$, can be determined. These $\theta_{max}$ values are listed in table I for the four contacts assuming the bulk Δ value. We note that the value for $\theta_{max}$ for each contact is consistently lower than θ obtained from the $d^2I/dV^2$ but given the limitation of the method in its present form, we find that the agreement is sufficiently convincing, to support our conclusions regarding the spin mixing angle in these films.

VI. SUMMARY

In this paper we have examined Andreev point contact spectra taken from $La_{1.15}Sr_{1.85}Mn_2O_7$ single crystal – Pb junctions, together with a compilation of literature Andreev point contact data on other highly spin polarised systems. We show that the existence of the Andreev Bound State resolves a number of long standing puzzles related to states with apparent and unphysical perfect transparency and high non-thermal broadening. We demonstrate that the zero bias conductance, the excess current and the position in energy of the sub-gap states can be used as a set of independent checks to quantify the degree of spin mixing at the SF interface. Taken together these results suggest that the study of single SF junctions can provide a robust method to gain valuable insight into the formation of ABS, the spin mixing angle and the quantification of factors that lead to the formation of the LRSTPE. Within the limits of the data presented, much of which have been re-examined from the literature, the average spin mixing angle appears to take a rather narrow range of values in highly spin-polarised materials, clustering around $0.5\pi$. The ferromagnets we have chosen to study share a common property in terms of their spin arrangement, and our observations may point to the importance of this property in determining the spin mixing angle in high transparency Andreev point contact spectra. This is in contrast to previous work examining the low transparency tunnelling regime using ferromagnetic Fe which has only partially spin polarised carriers and where it was found that the interfacial properties played a dominant role [44]. An important extension of our work would be to use the PCAR methodology in combination with external parameters that would allow direct variation of $\theta$ in-situ. Theoretically, using microwaves to excite ferromagnetic resonance (FMR) in the ferromagnetic layer should provide exactly such an external tuning of $\theta$ [70,71]. Andreev reflection could then be used to measure a single contact repeatedly as a function of microwave excitation frequency. This would enable a systematic variation of the spin mixing angle, and potentially changing the strength of the LRSTPE in SFS junctions using an external parameter. We anticipate that this will be an important development in future work.

**Acknowledgements**

J. W. A. R. acknowledges financial support from the Royal Society, the Leverhulme Trust through an International Network Grant (No. IN-2013-033), and the EPSRC Programme Grant EP/N017242/1. LFC acknowledges support from EPSRC grants EP/J014699 and EP/H040048.

Table I

List of parameters for each contact

| Contact name | Resistance at 10mV ($\Omega$) | Z (M-BTK) | Polarisation (M-BTK, %) | $\varepsilon/\Delta$ | $\theta$ from SGS | $\theta_{max}$ from IV assuming $\Delta$ = 1.35 meV |
|---|---|---|---|---|---|---|
| C, K | 66.7 | 0.08 | 81.5 | 0.5±0.07 | (0.66± 0.08) $\pi$ | ≤ 0.5 $\pi$ |
| E, E | 104 | 0 | 87.5 | 0.67±0.15 | (0.5±0.2)$\pi$ | - |
| E, G | 61.4 | 0.68 | 83 ± 3.5 | 0.45±0.10 | (0.7±0.1) $\pi$ | ≤ 0.4 $\pi$ |
| E, H | 58.7 | 0.63 | 86 ± 3 | 0.17±0.03<br>0.74±0.15 | (0.89±0.04) $\pi$<br>(0.39±0.24) $\pi$ | ≤ 0.5$\pi$ |

Figures:

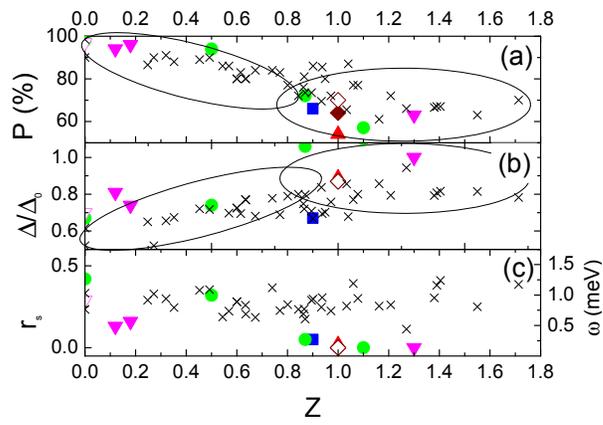

Fig. 1: Parameters as a function of *Z* extracted via the M-BTK model for Pb contacts on CrO$_2$ thin films (X) from Ref. [45] and from the literature from reference (■) [1], (▲) [72], (  ) [14],(▲) [7], (  ) [51], (◆) [73], (◆) [74], parameters taken from . Löfwander et al., ref [38]. (a) *P*, (b) Δ (c) broadening defined by a spreading resistance [38] or a thermal broadening parameter [45].

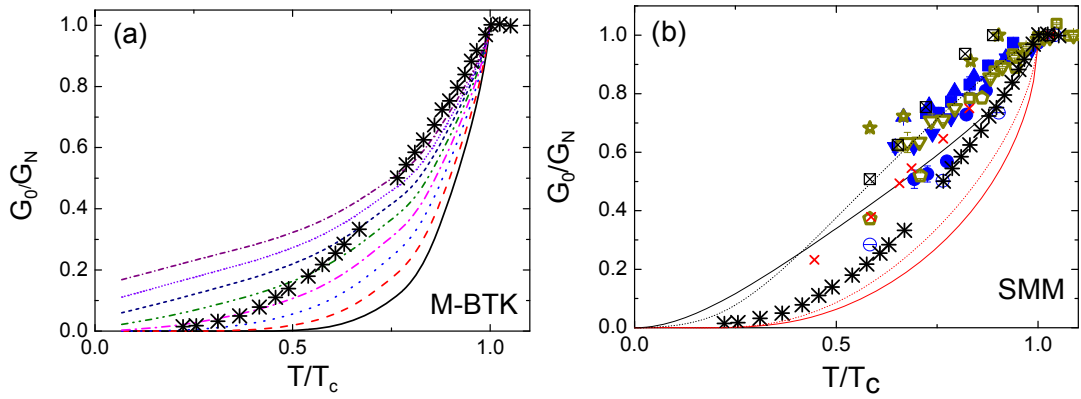

Fig. 2: (a) Temperature dependence of $G_0/G_n$ for the M-BTK model with thermal broadening only (solid black line) and then 0.1meV additional non-thermal broadening up to 0.7 meV additional non-thermal broadening. (*) $HgCr_2Se_4$ data taken from ref. [65]. (b) $G_0/G_n(T)$ for the SMM, with $\theta = 0$, $Z = 0.1$ (lower curve - solid red line), 0.5 (dotted red line), $\theta = \pi/2$, $Z = 0.1$ (upper curve – solid black line), 0.5 (dotted black line) with extracted data for $CrO_2$ films onto $TiO_2$ (blue symbols, [63], [15]), $Al_2O_3$ (Dark yellow, [45]), and $La_{1.15}Sr_{1.85}Mn_2O_7$ (x). Data extracted from literature data on LSMO3 (red, x) [7] and (*) on $HgCr_2Se_4$ [65].

Figure 3:

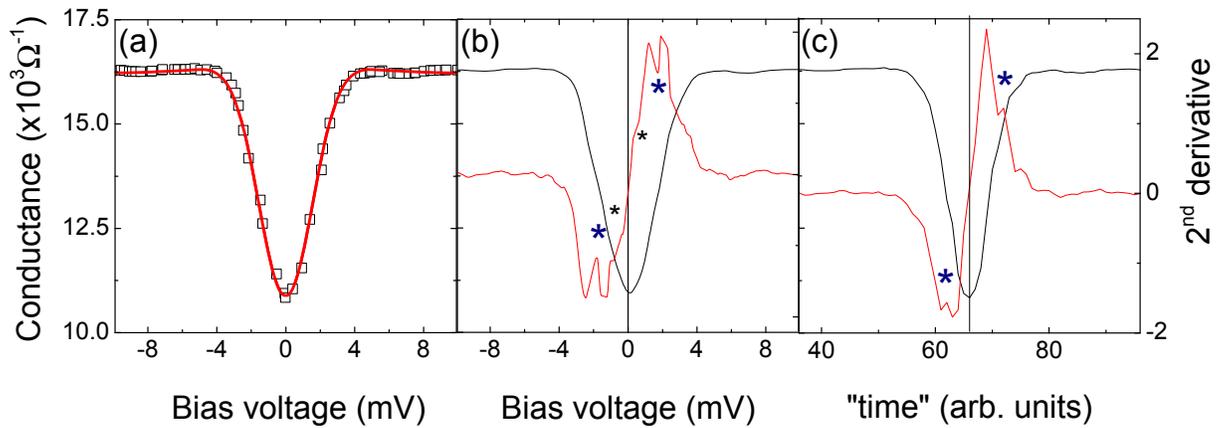

Fig. 3: (a) Conductance as a function of voltage for crystal E, contact G (squares) with the fit using the M-BTK model and P = 83%, Z = 0.68, Δ = 0.9 meV, ω = 1.05 meV. (b) Conductance for the same contact with the d²I/dV² showing peaks (in the negative bias) dips (in the positive bias) that may be associated with ABS. (c) conductance for the same contact as a function of acquisition order with the derivative of that data indicating that only the blue asterisked peak/dip feature is observed in both bias and time second derivatives.